# Application of Multilayer Feedforward Neural Networks in Predicting Tree Height and Forest Stock Volume of Chinese Fir

Xiaohui Huang[1], Xing Hu[1]
1 College of Software Engineering, Sichuan University, Chengdu, 610064, China
E-mail: hxhcindy@gmail.com
E-mail: helenahu6@gmail.com

Weichang Jiang[2], Zhi Yang[1], Hao Li[3*]
2 College of Computer Science, Sichuan University, Chengdu, 610064, China
3 College of Chemistry, Sichuan University, Chengdu, 610064, China
E-mail: dhcchemscu@gmail.com
E-mail: yzdedream@gmail.com
E-mail: lihao_chem_92@hotmail.com

*Abstract*-Wood increment is critical information in forestry management. Previous studies used mathematics models to describe complex growing pattern of forest stand, in order to determine the dynamic status of growing forest stand in multiple conditions. In our research, we aimed at studying non-linear relationships to establish precise and robust Artificial Neural Networks (ANN) models to predict the precise values of tree height and forest stock volume based on data of Chinese fir. Results show that Multilayer Feedforward Neural Networks with 4 nodes (MLFN-4) can predict the tree height with the lowest RMS error (1.77); Multilayer Feedforward Neural Networks with 7 nodes (MLFN-7) can predict the forest stock volume with the lowest RMS error (4.95). The training and testing process have proved that our models are precise and robust.

*Keywords*-Artificial neural networks, Multilayer Feedforward Neural Networks, Chinese fir, tree height, forest stock volume.

## I. INTRODUCTION

Wood increment, the increment of trees in a given period of time, is critical information in forestry management [1-2]. However, examining wood increment with a ruler manually consumes too much time and man effort. Therefore, it is meaningful to research for the precise, fast, economic and consistent method to measure wood increment. By establishing a model of forest stand, the dynamic growing status of the forest can be reflected objectively, and therefore the growth of forest stand can be predicted precisely. Previous studies used mathematics models to describe complex growing pattern of forest stand [3-5], in order to determine the dynamic status of growing forest stand in multiple conditions. In our research work, we aimed at studying the non-linear relationships of tree height and other properties, for establishing a precise and robust Artificial Neural Networks (ANN) model to predict the precise values of tree height and forest stock volume.

## II. MODEL DEVELOPMENT

### A. Fundamental of ANN models

Network model is widely used to describe the physical and biological phenomenon [6-8]. A neural network is composed of an interconnected group of artificial neurons,, and a connectionist way is taken to process information for it. In most circumstances, an artificial neural network (ANN) is an adaptive system that is equipped to be adapting continuously to new data and learning from the accumulated experience [9-10] and noisy data. Apart from that, the system structure can be changed based on external or internal information that flows through the network during the learning phase. Meanwhile, essential information can be abstract from data or model complex relationships between inputs and outputs.

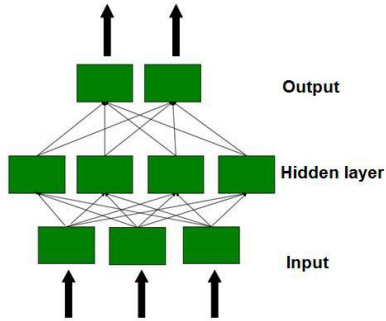

Figure 1. A schematic view of artificial neural network structure.

As can be seen from figure 1 above, the main structure of the artificial neural network (ANN) is made up of the input layer and the output layer. The input variables are introduced to the network by the input layer [11]. Also, the response variables with predictions, which stand for the output of the nodes in this certain layer, provided by the network .Additionally, the hidden layer is included. The type and the complexity of the process or experimentation usually iteratively determine the optimal number of the neurons in the hidden layers [12].

*B. Training process of ANN models*

The data of tree height and other forestry properties of Chinese fir are obtained from H.Q. Xie's research work [5] . In our studies, we chose time of life, crown density, slope, slope position and slope direction as the independent variable, while the tree height and forest stock volume as the dependent variables. The ANN models were developed by two kinds of typical models: General Regression Neural Networks (GRNN) [13-15] and Multilayer Feed-forward Neural Networks (MLFN) [16-18]. The nodes of MLFN models are set from 2 to 16, so that, we can observe the change of the RMS error. The results are the average of a series of repeated experiments, in order to ensure the robustness of the models. The training results of different ANN models are shown as table 2:

TABLE I. RESULTS OF DIFFERENT ANN MODELS IN PREDICTING TREE HEIGHT

| Model | RMS Error | Training Time | Reason Training Stopped |
|---|---|---|---|
| Linear Predictor | 32.30 | 0:00:00 | Auto-Stopped |
| GRNN | 2.69 | 0:00:00 | Auto-Stopped |
| MLFN 2 Nodes | 1.87 | 0:00:56 | Auto-Stopped |
| MLFN 3 Nodes | 6.15 | 0:00:46 | Auto-Stopped |
| MLFN 4 Nodes | 1.77 | 0:00:56 | Auto-Stopped |
| MLFN 5 Nodes | 4.05 | 0:01:15 | Auto-Stopped |
| MLFN 6 Nodes | 2.99 | 0:01:30 | Auto-Stopped |
| MLFN 7 Nodes | 3.59 | 0:01:58 | Auto-Stopped |
| MLFN 8 Nodes | 3.72 | 0:02:33 | Auto-Stopped |
| MLFN 9 Nodes | 1.32 | 0:04:12 | Auto-Stopped |
| MLFN 10 Nodes | 3.88 | 0:06:13 | Auto-Stopped |
| MLFN 11 Nodes | 4.56 | 0:11:46 | Auto-Stopped |
| MLFN 12 Nodes | 7.80 | 3:17:30 | Auto-Stopped |

Table 1 presents the results of different models in predicting tree height. We found that the MLFN model with 4 nodes (MLFN-4) and has the lowest RMS error (1.77). Therefore, we considered that the MLFN-4 model can generate the best model to predict the tree height of Chinese fir with the lowest time consuming and RMS error.

TABLE II. RESULTS OF DIFFERENT ANN MODELS IN PREDICTING FOREST STOCK VOLUME

| Model | RMS Error | Training Time | Reason Training Stopped |
|---|---|---|---|
| Linear Predictor | 5.39 | 0:00:00 | Auto-Stopped |
| GRNN | 7.32 | 0:00:00 | Auto-Stopped |
| MLFN 2 Nodes | 6.82 | 0:00:53 | Auto-Stopped |
| MLFN 3 Nodes | 7.10 | 0:01:06 | Auto-Stopped |
| MLFN 4 Nodes | 5.53 | 0:01:07 | Auto-Stopped |
| MLFN 5 Nodes | 8.74 | 0:01:34 | Auto-Stopped |
| MLFN 6 Nodes | 9.11 | 0:02:13 | Auto-Stopped |
| MLFN 7 Nodes | 4.95 | 0:02:10 | Auto-Stopped |
| MLFN 8 Nodes | 7.51 | 0:02:30 | Auto-Stopped |
| MLFN 9 Nodes | 10.57 | 0:04:24 | Auto-Stopped |
| MLFN 10 Nodes | 150.78 | 0:06:00 | Auto-Stopped |
| MLFN 11 Nodes | 9.00 | 0:12:34 | Auto-Stopped |
| MLFN 12 Nodes | 23.35 | 1:06:13 | Auto-Stopped |
| MLFN 13 Nodes | 9.20 | 7:35:35 | Auto-Stopped |
| MLFN 14 Nodes | 16.12 | 0:01:16 | Auto-Stopped |
| MLFN 15 Nodes | 30.70 | 0:01:47 | Stopped by User |

Table 2 presents the results of different models in predicting forest stock volume. We found that the MLFN model with 7 nodes (MLFN-7) and has the lowest RMS error (4.95). Therefore, we considered that the MLFN-7 model can generate the best model to predict the tree height of Chinese fir with the lowest time consuming and RMS error.

### III. RESULTS AND DISCUSSION

For more intuitionistic, six figures are used to describe the training results and testing results, which are shown as follows:

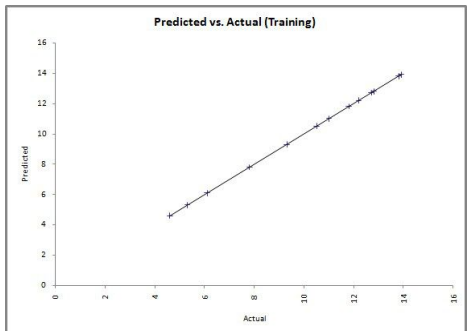

Figure 2: Comparison between predicted values and actual values during training process using MLFN-2 (tree height).

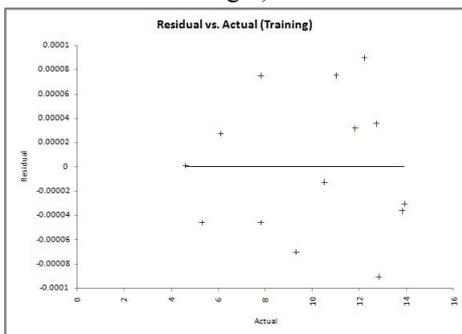

Figure 3: Comparison between residual values and actual values during training process using MLFN-2 (tree height).

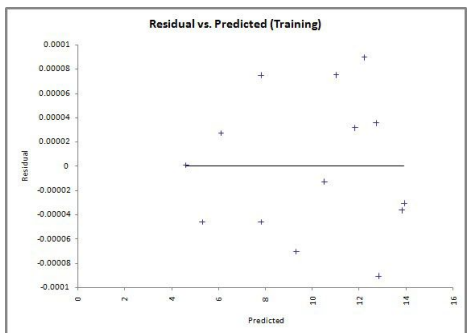

Figure 4: Comparison between residual values and predicted values during training process using MLFN-2 (tree height).

Figure 2 to 4 depict the results of training process in predicting tree height of Chinese fir, using MLFN-4 model. We found that the values are concentrated and corresponded with the normal training process of MLFN-4, showing that the training process is correct and precise.

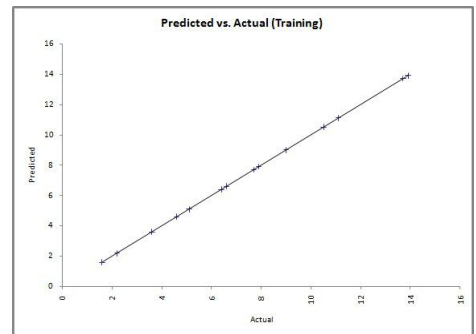

Figure 5: Comparison between predicted values and actual values during training process using MLFN-7 (forest stock volume).

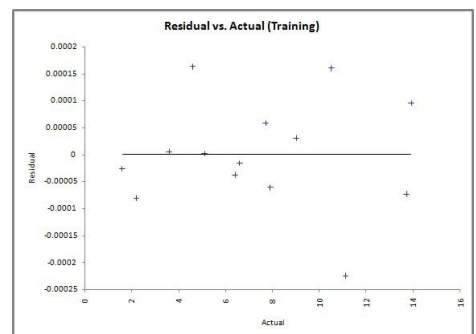

Figure 6: Comparison between residual values and actual values during training process using MLFN-7 (forest stock volume).

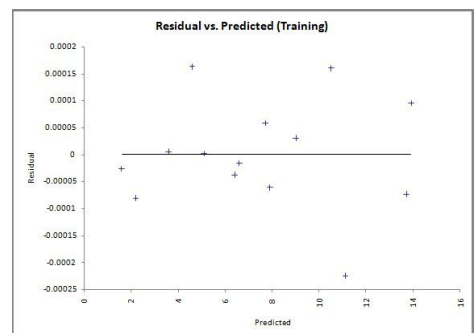

Figure 7: Comparison between residual values and predicted values during training process using MLFN-7 (forest stock volume).

Figure 5 to 7 depict the training process in predicting forest stock volume of Chinese fir, using MLFN-7 model. All the values shown in the three figures are the average values, from which we can draw a conclusion that the model is accurate and robust.

According to the training results, our models were proved to be suitable and reasonable in predicting the tree height and forest stock volume of Chinese fir.

IV. CONCLUSION

Wood increment is critical information in forestry management. Our research has successfully established Multilayer Feedforward Neural Networks (MLFN) models to describe the relationship between different properties of Chinese fir perfectly. Results show that Multilayer Feedforward Neural Networks with 4 nodes (MLFN-4)can predict the tree height with the lowest RMS error; Multilayer Feedforward Neural Networks with 7 nodes (MLFN-7) can predict the forest stock volume with the lowest RMS error. In future studies, we'll pay more attention to explore the relationship between different properties of forest stand.

V. ACKNOWLEDGEMENTS